\documentclass{aa}  

\usepackage{graphicx}
\usepackage{txfonts}
\usepackage{dcolumn}
\newcolumntype{m}[1]{D{,}{\,\pm\,}{#1}}

\usepackage[usenames,dvipsnames]{xcolor}
\usepackage{soul}

\usepackage[colorlinks = true,
linkcolor= blue,
citecolor = blue,
filecolor = blue,
urlcolor = blue,
]{hyperref}
\newcommand{\kms}{\,km\,s$^{-1}$} 

\begin{document}

   \title{ Kinematic origin of white dwarfs in the solar neighborhood}


   \author{Ainhoa Zubiaur
            \inst{1,2}
          \and
         Roberto Raddi\inst{1}\fnmsep\thanks{Email;
    roberto.raddi@upc.edu} 
          \and
           Santiago Torres\inst{1,3}}

\institute{Departament de F\'\i sica, 
           Universitat Polit\`ecnica de Catalunya, 
           c/Esteve Terrades 5, 
           08860 Castelldefels, 
           Spain
           \and
           Physikalisches Institut, Universit\"at Heidelberg, Im Neuenheimer Feld 226, 69120 Heidelberg, Germany
           \and
           Institute for Space Studies of Catalonia, 
           c/Gran Capit\`a 2--4, 
           Edif. Nexus 104, 
           08034 Barcelona, 
           Spain}

   \date{Received x, y; accepted x, y}

\titlerunning{Kinematic origin of the white dwarfs in the solar neighborhood}
\authorrunning{Zubiaur et al.}

\offprints{R. Raddi}

 
  \abstract
   {White dwarfs are considered to be efficient cosmochronometers.  Thanks to the recent space-borne mission {\it Gaia}, a nearly complete sample  up to about 100 parsecs from the Sun has been compiled. However, the Galaxy, as a dynamic system, implies that these objects may have very diverse origins. It is therefore of paramount importance to characterize the origins of white dwarfs from the different Galactic structure components found in our solar neighborhood.}
   {We aim to compute the Galactic orbits for white dwarfs of the thin and thick disk, as well as the halo components observed in our solar neighborhood. On the basis of these determinations, we  analyze the most probable regions of the Galaxy where they could have formed, along with the distribution of their orbital parameters and the observational biases introduced when constructing the local sample.}
   {We used a detailed Galactic orbit integration package, in conjunction with a detailed population synthesis code specifically designed to replicate the different Galactic components of the white dwarf population. Synthetic stars were generated based on the current observational sample and  their orbital integration allowed for the reconstruction of the population's history. }
   {Our kinematic analysis of the white dwarf population reveals the ephemeral nature of the concept of the solar neighborhood, as the majority of thin-disk, thick-disk, and halo white dwarfs will have left our 100\,pc neighborhood in approximately  3.30 Myr,  1.05 Myr, and 0.6 Myr, respectively. Moreover, the spatial distribution of the integrated thin-disk orbits suggests that 68\% of these stars were formed at less than 1 kpc from the Sun, while most of the thick-disk members have undergone radial disk migration. Halo members are those stars that typically belong to the ``inner halo,'' given that their orbits mostly planar and do not extend beyond $R = 20$--25 kpc. Despite the observational bias, which mostly affects the oldest stars in the thick disk and halo, we show that the wider distribution of orbital parameters is well represented by the  sample.}
   {The solar neighborhood is a transitory concept, whereby its current population of white dwarfs originates from larger regions of the Galaxy. This fact must be taken into account when analyzing the overall properties of such a population, such as its age distribution, metallicities, and formation history. Even so, the kinematic properties observed by recent missions such as {\it Gaia} are representative of the total population up to a distance of approximately 500 pc. 
   } 


   \maketitle
%

\section{Introduction}

The high-precision astrometry provided by the European Space Agency's \emph{Gaia} mission \citep{gaia} has opened up a wide parameter space to Galactic studies, enabling the mapping of stellar kinematics over a large volume of the Milky Way, extending up to 6 kpc towards the inner and outer Galactic radial directions \citep{gaia-kin}. The complementary information that is contributed by precise measurements of chemical abundances for ever-growing samples of stars is helping to shape the correlations between kinematics and stellar chemistry that are key to unraveling the past history of our Galaxy \citep{gaia-chem}. The improved astrometry delivered by the \emph{Gaia} Early Data Release 3 \citep[EDR3;][]{GaiaEDR3} has confirmed the interplay between the kinematics of local structures, such as moving groups \citep{antoja2008}, as well as that of more distant ones \citep{ramos2018,khanna2019} in the Galactic anti-center direction and away from the mid-plane of the Galactic disk. This implies a connection with the recent passing of the Sagittarius dwarf galaxy \citep{Ruiz-Lara2020} and interactions with the Large Magellanic Cloud \citep{antoja2018,laporte2018}.

The measurement of accurate stellar ages is crucial in this context, enabling to constrain the past history of the Milky Way that has led to the current spatial and velocity distributions of stars as well as their chemical compositions \citep[e.g.,][and references therein]{anders2023}. Stellar ages, along with masses, are the most difficult physical parameter to measure. However, some evolutionary phases are enough well understood that the modeling of individual objects can lead to reliable age determinations. One such phases is the white dwarf cooling sequence \citep[e.g.,][]{althaus10} that provides a powerful cosmochronological tool for constraining the age and the star formation history of the local disk and halo populations \citep[e.g.,][]{tremblay2014,kilic2017,torres2021}. The {\em Gaia} mission has led to the identification of large, statistically significant samples of white dwarfs \citep{Gentile-Fusillo2021}, which are effectively volume-complete within 100\,pc away from the Sun \citep{JE2018}, thus enabling an unbiased analysis of their population.

 Pioneering works focused on the kinematic analyses of white dwarf samples in the pre-\emph{Gaia} era \citep[e.g.,][]{silvestri2001,pauli2006,wegg2012,sion2014,anguiano2017}. However, more recently, by means of accurate \emph{Gaia} parallaxes and proper motions, it has been possible to associate white dwarfs to known stellar streams \citep{Torres2019a}, to measure the reflex solar motion \citep{rowell2019}, to assign their membership to Galactic components within 100\,pc \citep{Torres2019}, and to accurately measure their age-velocity dispersion relation \citep{cheng2019,Raddi2022}.
 
In this work, we model the kinematics of white dwarfs  with a synthetic population by integrating their Galactic orbits in a static potential and using the present-day boundary conditions that have been obtained from the \emph{Gaia} data. While a static potential is far from being a realistic approximation, the kinematics of local stars carries the imprint of resonances and radial mixing due to the Galactic structure and dynamics \citep{sellwood2002,dehnen2000,hunt2019}, thus enabling a statistical study of the distributions that characterize their orbital parameters. The crossing-time of white dwarfs that are passing through the solar neighborhood is numerically quantified, as it is known to introduce a bias, although small, in the average properties of classes of local stars \citep{mayor1977}. Moreover, we analyze the spatial distributions obtained by integrating for several orbits around the Galaxy as a statistical method for linking the observed white dwarf populations to the birth sites of their main sequence progenitors.

In the following sections, we introduce the properties of the observed and simulated samples (Sects.\,\ref{s:obsample} and \ref{s:sysample}, respectively), describe the orbit integration framework (Sect.\,\ref{s:orbitin}) and present a detailed analysis of the results (Sect.\,\ref{s:results}). In the last section, we conclude with a summary and the expectations for future developments.


\section{Observed sample}
\label{s:obsample}
The analysis of the white dwarf population provided by {\it Gaia} reveals that the   largest sample that can be considered as effectively complete is the 100 pc volume-limited sample \citep{JE2018}.  Thus, we  started from the sample analyzed in \citet{Torres2019}, in which a random forest (RF) algorithm was employed to classify the white dwarf population within 100 pc of the Sun into its Galactic components. The sample was selected by applying color-magnitude cuts to the  Hertzsprung–Russell diagram and adding  specific selection criteria, such as photometric and astrometric relative errors under 10 per cent \citep[see Sect. 4 from][for a detailed description]{Torres2019}.  Then we cross-matched it with the {\em Gaia} EDR3, providing updated astrometry for the same sample.

It is worth noting here that in order to eliminate possible contaminants in high-crowded areas, the observed sample analyzed in \citet{Torres2019} did not include objects with proper motions corresponding to projected tangential velocities below $\varv_{\rm tan} = 5$\,\kms.  This fact should be taken into account in our analysis because it may introduce a slight bias for low-speed objects that are absent from the observed sample.

The observed sample was thus classified using a RF algorithm applied to a 8D space
(equatorial coordinates, parallax, proper motion components, and photometric magnitudes).  The updated sample, with {\it Gaia} EDR3 astrometry, delivers 11,568 thin-disk, 1317 thick-disk, and 89 halo white dwarf candidates, implying a reduction of 6\,\% for each of the populations with respect to the number of candidates reported by \citet{Torres2019}. Such a small difference is attributed to stars that are now outside the 100\,pc volume. We note that we did not search for new 100-pc candidates in EDR3. Although radial velocity measurements are absent in the \citet{Torres2019} sample, it may be considered the largest, nearly complete, sample of white dwarfs, thus justifying a reliable analysis of the kinematics of the Galactic components of this population.

 Since most of the known white dwarfs do not have accurate radial velocity measurements, with very few of them having suitable high-resolution observations \citep{Napiwotzki2020}, a radial velocity of $\varv_{\rm rad}$ = 0\,\kms has been usually adopted in kinematic studies of this population.  Alternative methods to overcome this issue have been proposed \citep{Dehnen1998}; however, they are not exempt from certain limitations \citep{McMillan2009}.  In order to see what is the effect of considering the radial velocity zero, \citet{Torres2019}, analyzed the velocity component distributions in the Toomre diagram for the synthetic sample of the different Galactic components by subtracting the $\varv_{\rm rad}$. Their analysis showed an average bias of 10\,\kms\ and up to 74\,\kms\ for the average $V$ component of the disk and halo stars, respectively, when the $\varv_{\rm rad} = 0$\,\kms. The bias resulting by the assumption of zero radial velocity for white dwarfs is estimated to cause a reduction by a factor of one-third in the three velocity components of the heliocentric Cartesian reference system \citep[e.g.,][]{rowell2019}. Although it is beyond the scope of the present work to retrieve the full 3D velocity space of the white dwarf population, we address this issue in Section\,\ref{ss:radial}, where we attempt to correct for this bias using our synthetic model.

\section{Synthetic white dwarf population}
\label{s:sysample}

\begin{table*}[t]
\caption{Mean value and dispersion for the initial Gaussian velocity distributions, as well as the covariance between $U$ and $V$.}
\centering

\begin{tabular}{@{}l m{5.4} m{7.4} m{5.4} m{5.0}@{}}
\hline 
      \noalign{\smallskip}
Population & \multicolumn{1}{c}{$U$}  & \multicolumn{1}{c}{$V$ } & \multicolumn{1}{c}{$W$} & \multicolumn{1}{c}{ Cov[$U$,\,$V$]} \\ 
 & \multicolumn{1}{c}{[\kms]}  & \multicolumn{1}{c}{[\kms]} & \multicolumn{1}{c}{[\kms]} & \multicolumn{1}{c}{[km$^2$\,s$^{-2}$]} \\ 
      \noalign{\smallskip}
\hline 
      \noalign{\smallskip}
Thin disk &  -4.23 , \sigma_u (T_{*}) &  -14.67 , \sigma_v (T_{*}) &  -4.78 , \sigma_w (T_{*}) &  22\\
Thick disk &  -29.82 ,  59.94 &  -48.63 ,  34.69 &  -6.43 ,  39.37 &  -181\\
Halo &  -41.17 ,  120.46 &  -123.34,  78.29 &  -6.79 ,  78.31 &  -995 \\   
      \noalign{\smallskip}
\hline
\multicolumn{5}{l}{Notes: T$_{*}$ is the total age of the star, which is the sum of the cooling time}\\ 
\multicolumn{5}{l}{of the white dwarf and the age of the progenitor. The velocities are given}\\
\multicolumn{5}{l}{in the heliocenric reference system.}\\
\label{table:vel-gauss}
\end{tabular}
\end{table*}

In addition to the observational sample described above, we produced a synthetic sample of white dwarfs to address potential observational biases or deficiencies in the observational sample. We took advantage of the consolidated population synthesis code based on Monte Carlo (MC) techniques described in \citet{Torres2019}, which has been widely used over the last few decades in the study of the disk, halo, Galactic bulge, open clusters, and globular clusters \citep[e.g.,][]{GB1999,Torres2002,GB2010,Cojocaru2015,Torres2015,Torres2016,Torres2018} 

 The MC algorithm provides the initial position of each star within a 100\,pc radius around the Sun and its age. In the following, we provide some of the basic ingredients used in our population synthesis code. According to a certain star formation rate (SFR), a probabilistic distribution determines the the ages of stars belonging to each Galactic component. We assumed a uniform SFR for the thin disk that lasts for a lifetime of $T_{\rm age}$, along with Gaussian distributions for the thick disk $\mathcal{N}$($T_{\rm age}$,\,1\,Gyr) and the halo $\mathcal{N}$($T_{\rm age}$,\,0.5\,Gyr),  with $T_{\rm age} =$ 9.2, 10.0 and 12.5\,Gyr, respectively \citep{Torres2019}.

The local standard of rest (LSR) is the chosen reference frame to describe the positions and velocities of the white dwarfs. It is instantaneously located on the Sun at $R_0 = 8$\,kpc from the Galactic center, performing a circular orbit rotating at $V_0 = 220$\,\kms, which is nearly the average rotation velocity of stars in the solar neighborhood \citep[][]{bovy2012}. We placed the Sun at $z = 25$\,pc above the Galactic plane \citep[e.g.,][]{juric2008}. Nevertheless, the actual values for the Sun's location within the Milky Way and the rotation speed at the Galactocentric distance of the Sun are irrelevant because we are considering a local population, thus,  our simulation is scalable to more recent estimates \citep[e.g.,][]{gravity2019}.

The simulated white dwarf spatial distribution follows double-decreasing exponential profiles in the Galactocentric cylindrical coordinates, $R$ and $z$, with given scale length and height, $L$ and $H$, for the disk components. We adopted  $H =$ 250\,pc and $L =$ 2.6\,kpc for the thin disk and $H =$ 1.5\,kpc and $L =$ 3.5\,kpc for the thick disk \citep{Torres2019}. An isothermal distribution, with a local constant density, was assumed for the halo members.  

Regarding the velocity distribution, we adopted a Schwarzschild distribution where each Galactic component $(U, V, W)$ follows a Gaussian distribution that reproduces the average quantities observed for white dwarfs  with respect to the heliocentric reference frame (Table \ref{table:vel-gauss}). That is to say, the adopted mean values are those corresponding to the observed sample  following the classification provided by \citet{Torres2019}, with the astrometry updated to {\em Gaia} EDR3, while the assigned velocity dispersion of the thin-disk stars is a function of their age, for which we adopted the average values of the full sample studied by \citet{Raddi2022} (see their Table 8). As the adopted values were drawn from observed samples,  they naturally account for the asymmetric drift that is typical of each given sub-population.  We note that our synthetic population accounts for the vertex deviation, namely, the correlation between $U$ and $V$ velocity components. The covariance of the $UV$-components for the observed stars are listed in Table\,\ref{table:vel-gauss}.  From the observed sample, we measured small correlations ($r_{\rm thin} = 0.04$, $r_{\rm thin} = -0.09$, $r_{\rm halo} = -0.1$), which  correspond to a deviation of less than $\pm 6^{\circ}$ in the UV plane (positive for the thin disk, negative for the thich disk and halo). A larger value of around $15^{\circ}$ was found by \citet{rowell2019}; however, this value is based on a white dwarf sample extending up to 250\,pc. Moreover, when modeling the kinematic of the 100-pc sample, we did not account for purely local peculiar velocity groups, such as the Hercules stream, which was shown to cause a strong overdensity of white dwarfs belonging to the thick disk in the $UV$-plane that are lagging behind the LSR \citep{Torres2019a}. By not accounting for the Hercules stream, it is expected that our simulated thick disk population will have a slower average velocity with respect to the observed population.

Finally, we also note that our halo sample is centered on a much smaller value for $V$ with respect to \citet{kilic2019}, but only 55 out of 142 halo candidates that were identified by those authors as 5-$\sigma$ disk outliers are found within 100\,pc from the Sun. Thus their selection is implying a likely  extreme kinematics that is compatible with the net zero rotation of the halo, as observed by larger surveys that target more distant stars, while the 100-pc sample will contain more inner halo objects that can possess an average rotation.

\section{Orbit integration}
\label{s:orbitin}

The orbit integration has been carried out by means of the  \verb|python| library  \verb|galpy| \citep{Bovy2015}.  The integration for each white dwarf orbit is initialized with a chosen Galactic potential, an array of time-steps and an array of 
the form \verb|[R,vR,vT,z,vz,phi]|, being \verb|(R,phi,z)| the Galactocentric radius, azimuth, and height above the plane, respectively, in the cylindrical system and \verb|(vR,vT,vZ)| the Galactocentric velocity components that are equivalent to the $(U, V, W)$ in the Cartesian reference frame. The adopted peculiar motion of the Sun with respect to the LSR is $U_\sun = \,-11.1$\,\kms, $V_\sun =$\,12.24\,\kms, and $W_\sun =$\,7.25\,\kms\ according to \citet{Schonrich2010}. The reference system of \verb|galpy| is left-handed, with the positive $R$ axis pointing away from the Galactic center towards the Sun and the positive $V$ component along the direction of the Galactic rotation.

We used the standard Galactic potential, \verb|MWPotential2014| \citep{Bovy2013,Bovy2015} for the orbit integration of both the observed and simulated populations. This is a static, axisymmetric potential, which does not include perturbative effects due to the presence of the Galactic bar and spiral arms. It consists of three time-independent components, whose relative amplitudes are scaled to match observed properties of the Milky Way \citep{Bovy2015}: a spherical bulge modeled as a power-law density profile with an exponential cutoff \citep{Bovy2015}, an axisymmetric disk \citep{Miyamoto1975}, and a dark-matter halo \citep{Navarro1997}. We note that the bulge has a lower mass with respect to values given in other works \citep{Bovy2013}, implying reduced gravitational forces for objects that eventual approach the Galactic center. Further details on the scaling of the relative Galactic components are presented in \citet{Bovy2015}.

Orbits were integrated in a time lapse of 2\,Gyr with $1\,000$ time-steps. This period corresponds to almost nine complete orbits of the Sun around the Galactic center (about 224\,Myr), ensuring the statistical significance of our analysis. Furthermore, in the analysis of the solar neighborhood's lifetime (Section\,\ref{ss:lifetime}), we performed a shorter integration for 15\,Myr with 1\,000 time-steps.

\section{Results}
\label{s:results}

The solar neighborhood within the Galaxy can be defined as the region of space that is enclosed by a certain radius from the Sun, in our case 100\,pc. Therefore, a well-defined solar neighborhood is an essential concept because a volume-limited sample from this region can be later used to extrapolate the local properties of a given stellar population to the rest of the Galaxy.

As the Sun orbits the Galactic center, the solar neighborhood also moves. In other words, we have adopted a Lagrangian definition for our neighborhood space. Therefore, we expect to have a flux of objects entering and exiting this volume. We emphasize that the results discussed in this section are valid on timescales shorter than the Milky Way's lifetime; hence, they describe the average properties of a white dwarf sample in the solar neighborhood.

\subsection{Orbital analysis of the solar neighborhood}

\subsubsection{Average solar neighborhood lifetime}
\label{ss:lifetime}
\begin{figure}[t]
    \centering
    \includegraphics[scale = 0.15]{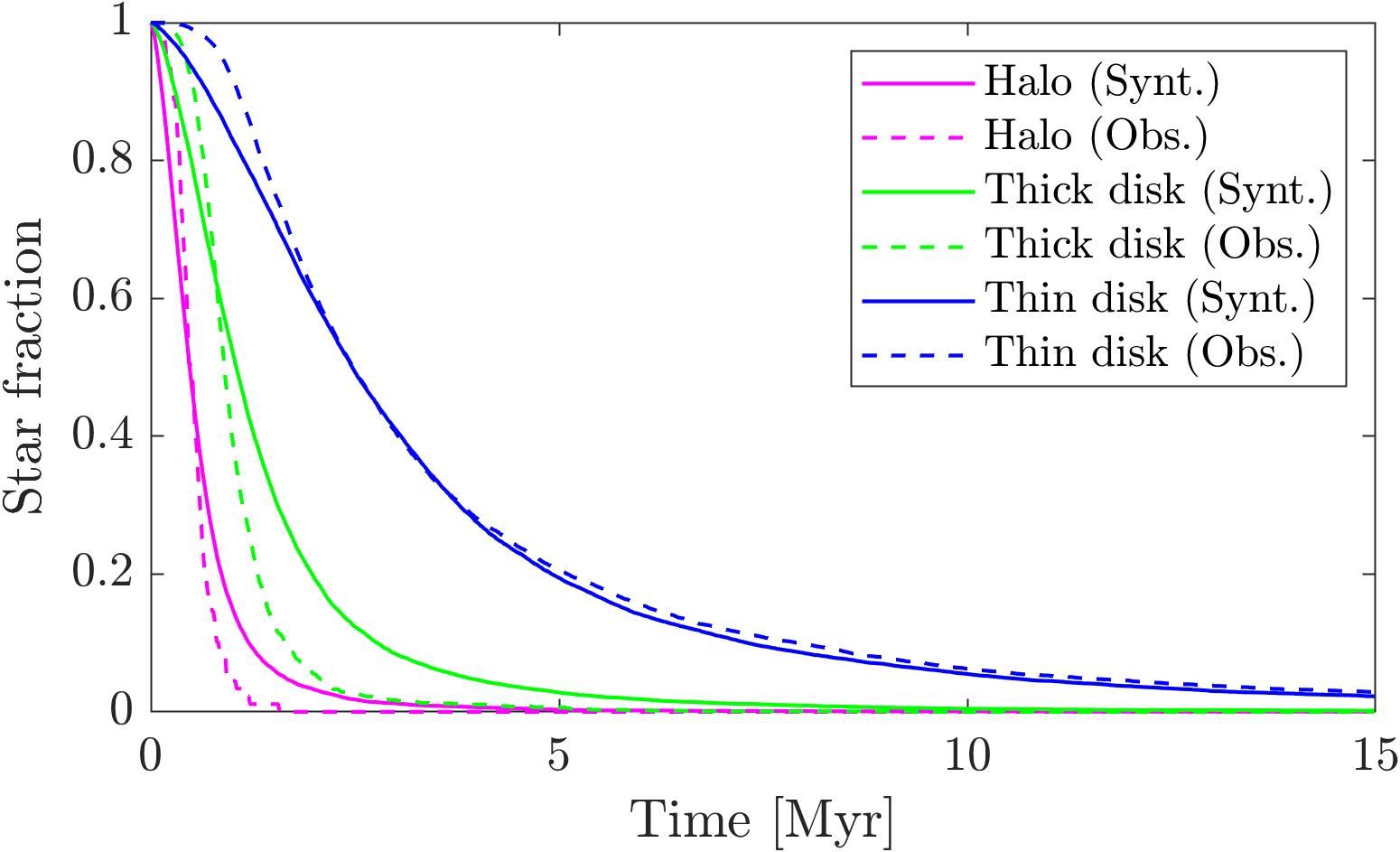}
    \caption{Fraction of observed and simulated stars that remain within a 100\,pc distance from the Sun for each Galactic component.}
    \label{fig:sdisp}
\end{figure}

In this first exercise, we counted the stars that exit the solar neighborhood as a function of time. The goal is to measure what is the rate of ``renewal'' of white dwarfs within this volume. Figure \ref{fig:sdisp} shows the results for our two reference samples of white dwarfs, the {\it Gaia} observed  sample (dashed lines) and the synthetic one (continuous lines; see Sections \ref{s:obsample} and \ref{s:sysample}, respectively), color-coded according to the different Galactic components considered.

The analysis of the results clearly reveals that the fraction of the initial number of stars ($y$-axis of Fig. \ref{fig:sdisp}) rapidly decreases. In less than 10\,Myr, more than 90\% of the initial number of white dwarfs have left our solar neighborhood. This fact can be  quantified by assuming an exponential decay and  computing then the lifetime, namely, the elapsed time for the star fraction to decrease by an $e$ factor. The results are given in Table \ref{table:lifetime} for the different populations considered. As expected, the dispersion effect is more pronounced for thick disk objects than thin-disk ones and even more for halo objects that possess, on average, larger velocity dispersion relative to the Sun. In particular, in approximately  3.3\,Myr the majority of thin-disk white dwarfs have left our solar neighborhood, while the thick disk stars are doing it in  $\sim 1.4\,$Myr and the halo stars scatter off in less than 1\,Myr.

\begin{table}[t]
\caption{Mean lifetimes of the solar neighborhood for each Galactic component.}
\label{table:lifetime}
\centering                         
\begin{tabular}{@{}lcc@{}}
\hline 
      \noalign{\smallskip}
& Synthetic sample & Observed sample \\ 
      \noalign{\smallskip}
\hline 
      \noalign{\smallskip}
Thin disk &  3.30\,Myr  &  3.27\,Myr  \\
Thick disk &  1.38\,Myr &  1.05\,Myr \\
Halo &  0.63\,Myr &  0.60\,Myr \\
      \noalign{\smallskip}
\hline     
\end{tabular}
\end{table}

It is also worth noting that the observed  thick disk sample decays  30\,\% faster than the corresponding synthetic equivalent sample. That would indicate, on average, faster objects in the observed sample than in the synthetic one. This fact can be  mainly attributed to 
 our synthetic model's  omission of the kinematics of  stellar streams, such as the Hercules stream. This stream, observed in white dwarf kinematics \citep{Torres2019a},  typically lags behind the average motion in the $UV$-plane and it is due to a resonance with the Galactic bar \citep[e.g.][]{antoja2008}. The net effect is a slight increase in the average values of the velocity components in the Galactic plane and a deviation from the Gaussian distribution of velocities for the observed sample. Effects that would contribute to the discrepancy found in Table \ref{table:lifetime} with respect to the synthetic model.

 We also stress that the exclusion or inclusion of low-speed objects ($< 5$\,\kms; see Section\,\ref{s:obsample}) from the observed and simulated samples does not change the results by more than 1\,\%.   Moreover, we note that the candidates in the observed sample have been classified using a RF algorithm, which tends to under represent low-speed objects for the Galactic thick disk and halo  components by classifying them as thin-disk stars. Consequently, both factors tend to increase the average speed of the different Galactic components  of the observed sample.

We conclude that the mean lifetime of the different white dwarf populations is very short compared to the orbital period of the Sun around the Galaxy. Thus, the concept of white dwarf solar neighborhood  should be understood as an ephemeral concept, as there are no physical processes in the lives of these stars, except for the planetary nebula phase, with a duration shorter than or on the order of a couple of million years. Nevertheless, in taking into account the long evolutionary timescales of white dwarfs and assuming a space density that is constant at low Galactic latitudes, the population that is currently residing in the solar neighborhood is assumed to be representative for statistical studies (see Sects. \ref{ss:size} and \ref{ss:selection}).

\subsubsection{White dwarf birth places}
\label{ss:wdplaces}

\begin{figure}
    \centering
    \includegraphics[width=0.85\columnwidth]{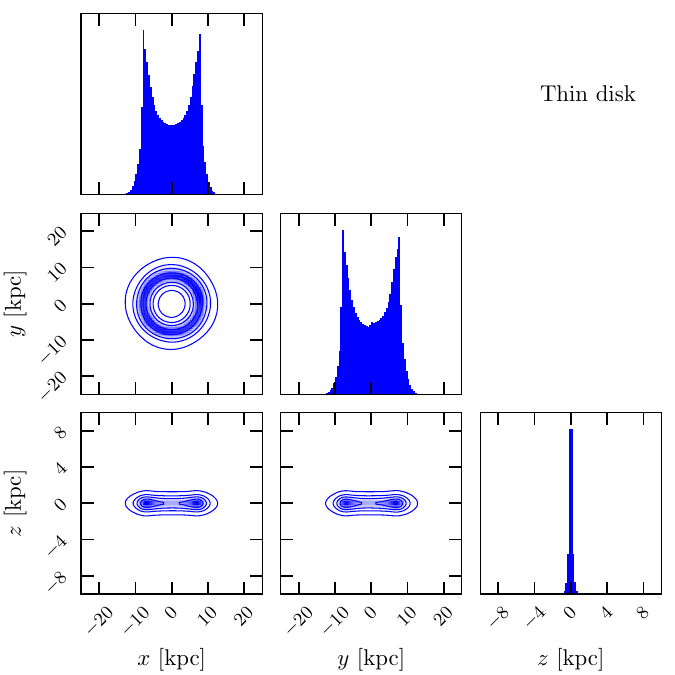}
    \includegraphics[width=0.85\columnwidth]{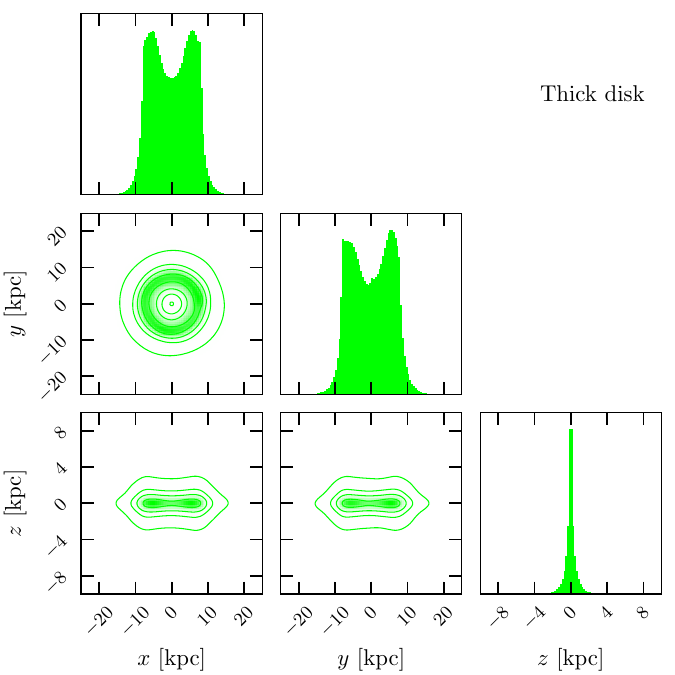}
    \includegraphics[width=0.85\columnwidth]{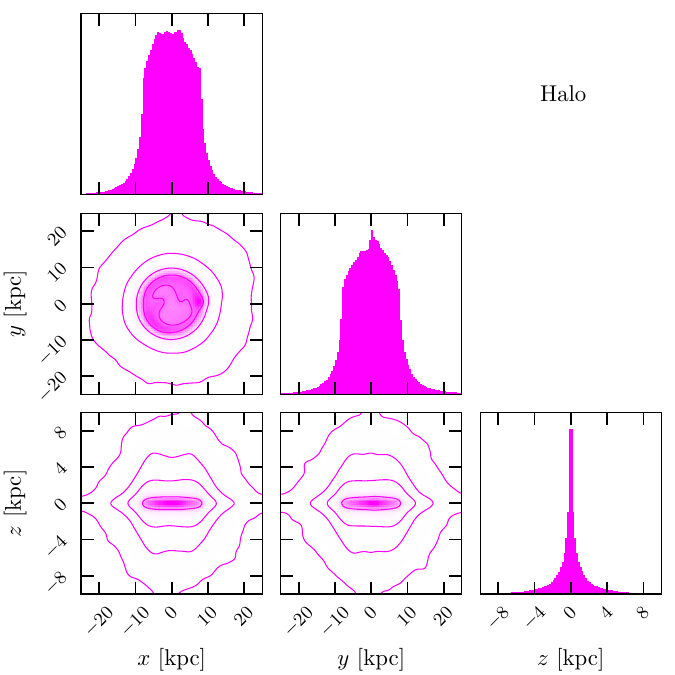}
    \caption{Corner plots showing the spatial distribution of integrated orbits of the simulated populations in Galactocentric Cartesian coordinates; from top to bottom, we display the thin-disk, thick-disk,  and halo samples, respectively. The density contours delimit the 1, 3, and 5\,$\sigma$ levels and the 99th percentile of the 2D distributions.}
    \label{fig:corn}
\end{figure}

Identifying the exact coordinates where individual stars come from is indeed a futile endeavor for objects as old as the white dwarfs. However, we can study the regions of the Milky Way where each population primarily resides and use this information as a proxy for their most likely birth places. In order to do this, we computed the orbits for nearly $15\,000$ white dwarf for each Galactic component and adopted a 2\,Gyr time interval to integrate the orbits of the synthetic sample. Such a long time interval (in comparison with the dissipative time previously estimated) ensures a full description of the parameter space that is covered by each star of our sample.

In Fig. \ref{fig:corn}, we show the spatial Galactocentric distributions $(x, y, z)$ covered by the orbits of the thin and thick disk, and halo populations. For each Galactic component, we represent them in a corner plot (where the top panels correspond to histogram distributions, while the rest of the panels show the density plots of the different coordinate projections). As expected, halo members (bottom corner plot) show a wider spatial distribution within our Galaxy than the disk components, while thin-disk (top corner plot) objects are concentrated in a more defined ring in the $xy$ plane, as compared to the thick-disk population (middle corner plot).

A closer look at the top corner panel in Fig. \ref{fig:corn} reveals the small deviation the thin-disk members have with respect to an almost-circular orbit, which are due to their small velocity dispersion. The $xy$ projection shows a truly thin ring, with a $2$\,kpc width at the $1\sigma$ level, which is indicative of the fact that the orbits are mostly circular.   There is also a narrow span in $z$ in which they merely deviate from the disk plane ($\pm$ 330\,pc also at the $1\sigma$ level), reaching 1\,kpc  at most in a few extreme cases.

With respect to the thick-disk population (middle corner plot), the results reveals a wider spatial distribution of stars, reaching larger values in $z$ ($\pm$ 780\,pc at  $1\sigma$). Meanwhile, the maximum deviations from the solar location in $x$ and $y$ are similar to those of the thin disk, the deviations in $z$ are almost the double. The “ring” in the $(x, y)$ projection of thick disk orbits broadens, reaching  3.4\,kpc, suggesting more eccentric orbits associated to the members of the thick disk population.

Finally, the halo population represented in the bottom corner plot of Fig. \ref{fig:corn} shows more distinct characteristics than the two previous disk populations. The two peaks presented in the $x$ and $y$ distribution for the disk members seem to merge in the case of the halo. This means that the orbits of halo objects are distributed in a more uniform way than those of disk members. Besides, the orbits of halo white dwarfs have a significantly larger extent in all directions, occasionally exceeding the radius of the Sun and moving several kpc above the Galactic plane. The mean deviation of halo orbits with respect to the plane is  2.5\,kpc and their radial extent is of  $\approx  6.7$\,kpc.

As a complement to the spatial distribution discussed, we computed the guiding-center radius, $R_g$, defined as the radius that a circular orbit would have for a given $z$ component of the orbital angular momentum. The guiding-center radius assumes the validity of the epicyclic approximation for the stellar motion in the Galactic potential and its a good proxy for the mean orbital motion of a given star \citep{kordopatis2014}. It is suggested that the effects of increased eccentricity and radial migrations due to interactions at the Lindblad resonances and the corotation radius in the Galactic disk \citep[``blurring;''][]{sellwood2002,schonrich2009} can be removed by investigating the properties of stars as a function of their guiding-center radii \citep{kordopatis2014}. One way to calculate this value is by means of a \verb|galpy| function defined as $R_g = R \times \frac{V}{V_0}$, where $R$ is the Galactocentric radius, and $V$ is the circular velocity at the solar radius (i.e., the $V$ component in the Cartesian reference frame).

\begin{figure}
    \centering
    \includegraphics[scale = 0.15]{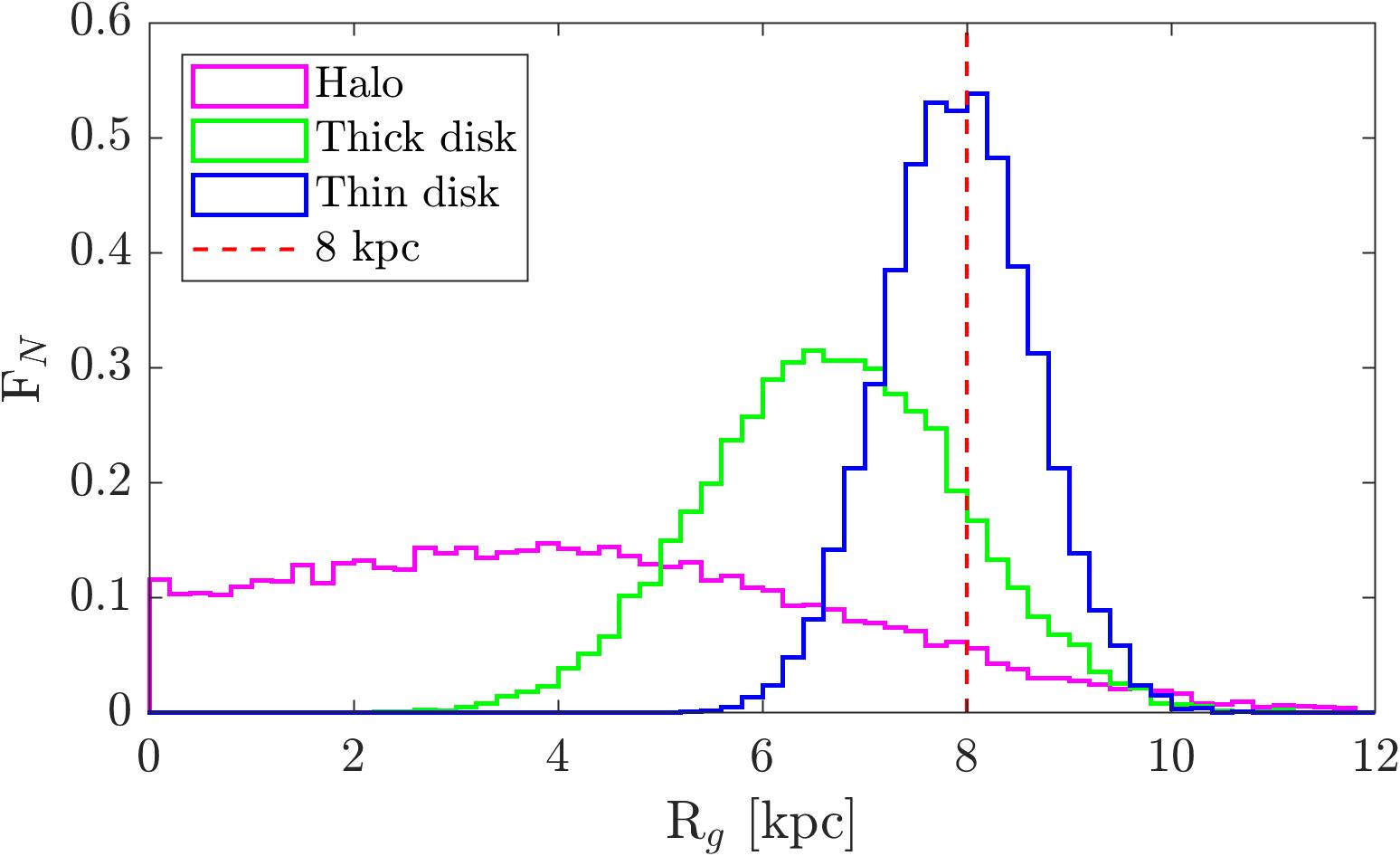}
    \caption{Area-normalized distribution of guiding radii for the different Galactic components considered in this study.}
    \label{fig:Rg}
\end{figure}

In Fig. \ref{fig:Rg}, we represent the distribution of $R_{\rm g}$ for the different Galactic components of our synthetic population. As a visual aid, we represent the position of the Sun ($R_0=8\,$kpc) by a vertical red dashed line. Table \ref{table:Rg} lists the mean values and standard deviations for these distributions. 

The guiding-center radius distribution of the thin-disk population is, as expected, centered at the Sun's location with a relative small dispersion of values both towards and outwards the Galactic center. This result suggests that 68\% of stars in the solar neighborhood where formed at less than 1\,kpc away from the Sun. We note a similarly symmetric distribution for the thick disk population, although centered at a smaller radius, implying that those objects with $R_{\rm g}$ suffered with radial migration of their average orbits. Finally, the halo population shows a much wider distribution than the disk populations. Even more, halo stars spend longer time closer to the Galactic center than their current position,  namely,  92.16\,\% of them have a $R_{\rm g}<8$\,kpc. Thus, the halo population that we currently see in the solar neighborhood belongs to what is likely the stellar components of the ``inner halo,'' given that their orbits are mostly planar and do not extend beyond $R = 20$--25 kpc. The broad range of $R_{\rm g}$ that the synthetic population present is due to the larger Galactic orbit eccentricities that are measured for the observed halo stars \citep[e.g.,][]{Torres2019,Raddi2022}. Their immersion into the inner regions of the Milky Way is also indicative of stronger interactions with the Galactic bar and spiral arms, which may have scattered them from initially circular disk orbits. 

\begin{table}
\caption{Mean values and standard deviations for the guiding-center radius distributions of each galactic component.}
\label{table:Rg}
\centering                         
\begin{tabular}{@{}lcc@{}}
\hline 
      \noalign{\smallskip}
& $\langle$R$_g$ $\rangle$\,[kpc] & $\sigma_{R_g}$\,[kpc] \\ 
      \noalign{\smallskip}
\hline 
      \noalign{\smallskip}
Thin disk & 7.9 & 0.7 \\
Thick disk &  6.7 &  1.3 \\
Halo &  4.2 &  2.5 \\
      \noalign{\smallskip}
\hline     
\end{tabular}
\end{table}

\subsubsection{Distribution of orbital parameters}
\label{ss:orbits}

\begin{figure}
    \centering
    \includegraphics{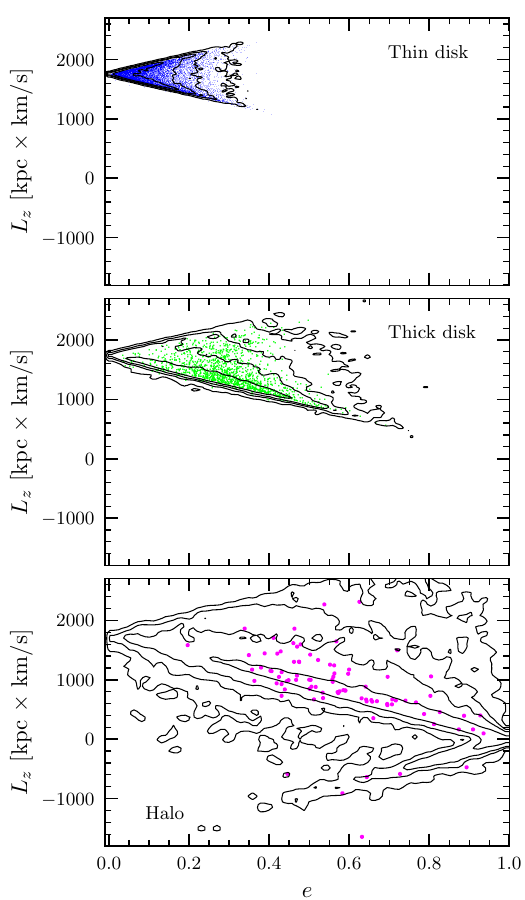}
    \caption{$Z-$component of the angular momentum vs eccentricity. Contours represent the 1, 3, and 5\,$\sigma,$ along with the 99th percentiles of the simulated distributions, while the scatter points represent the observed white dwarfs within 100\,pc \citep{Torres2019}.}
    \label{fig:ecc_Lz}
\end{figure}

The orbit integration with \verb|galpy| allows us to obtain the integrals of motion and various orbital parameters for each trajectory of individual stars. We first explore the distribution of the angular momentum perpendicular to the Galactic plane, $L_z$, as a function of the eccentricity of the orbit, $e$, for the objects of each Galactic component. The results are shown in Fig. \ref{fig:ecc_Lz} as density contour lines for the synthetic population, and scatter dots for the observed sample.

\begin{figure}
    \centering
    \includegraphics[scale = 0.15]{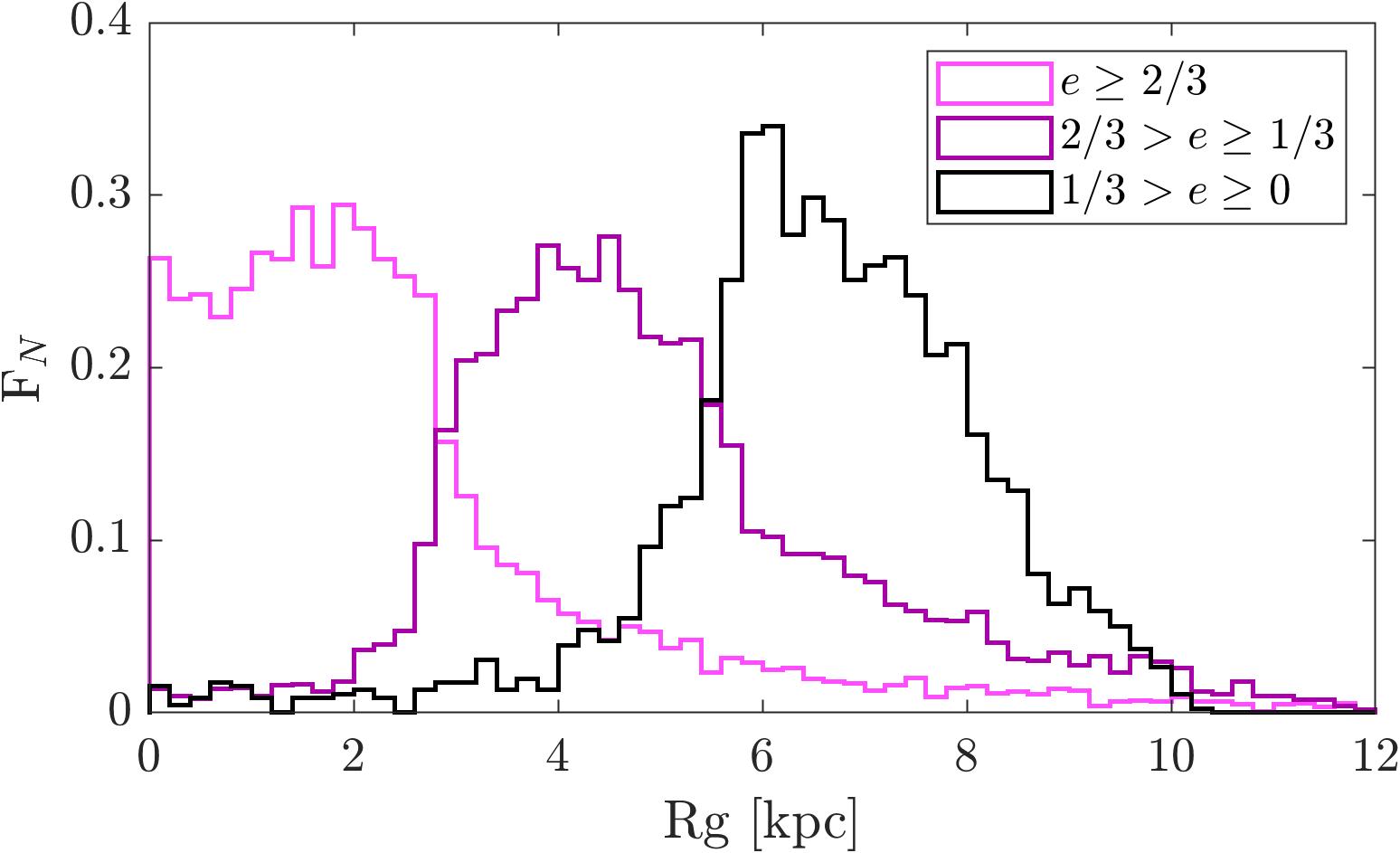}

    \includegraphics[scale = 0.15]{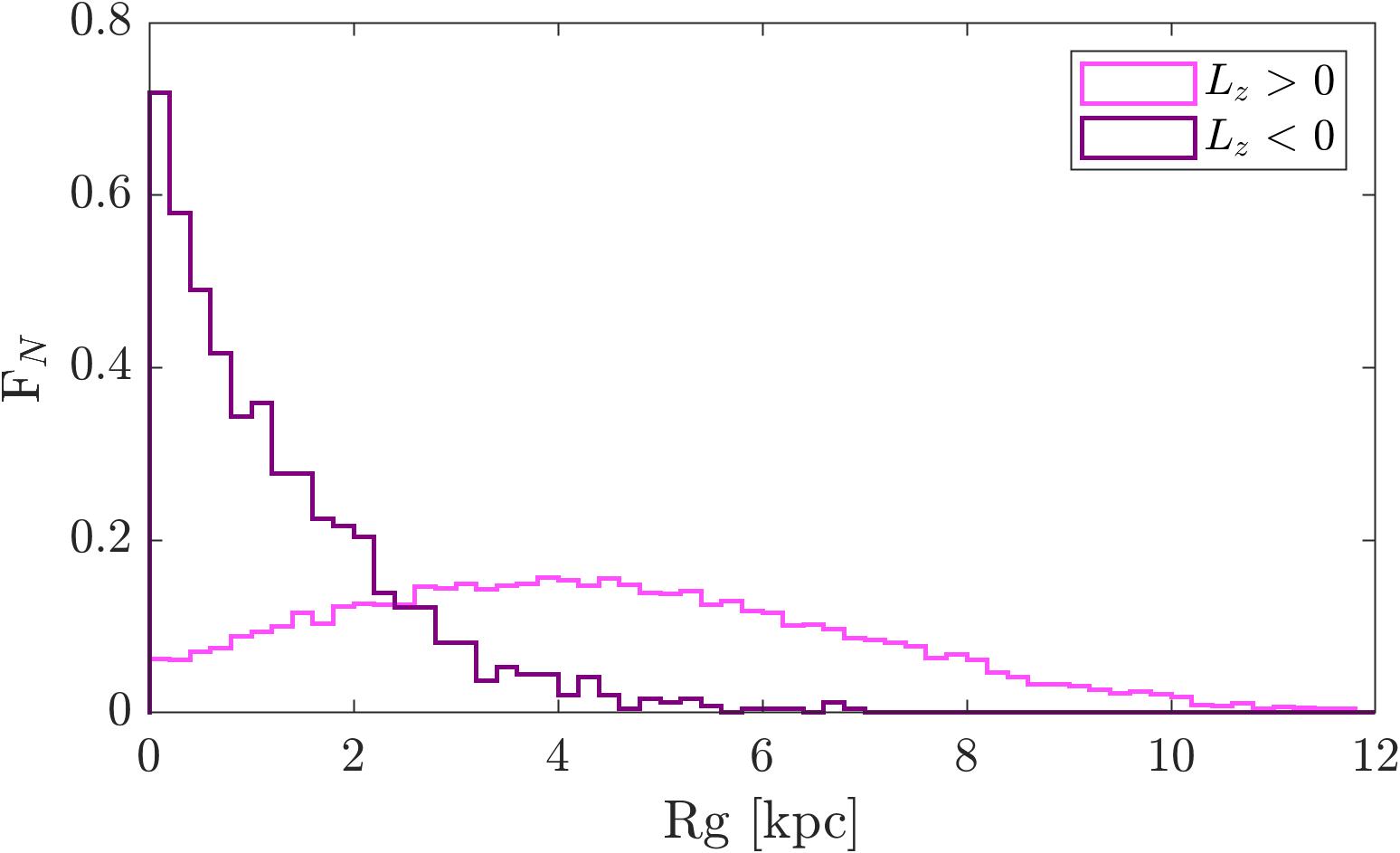}
    \caption{Distribution of guiding-center radii for the simulated halo white dwarfs, divided in different ranges of eccentricity (top panel: $e \in (0,1/3)$, $e \in (1/3,2/3),$ and $e \in (2/3,1)$) and $z$-component of the orbital angular momentum (bottom panel: $L_z$ > 0 and $L_z$ < 0 in the bottom panel).}
    \label{fig:Rg_e_Lz}
\end{figure}
\begin{figure}
    \centering
    \includegraphics{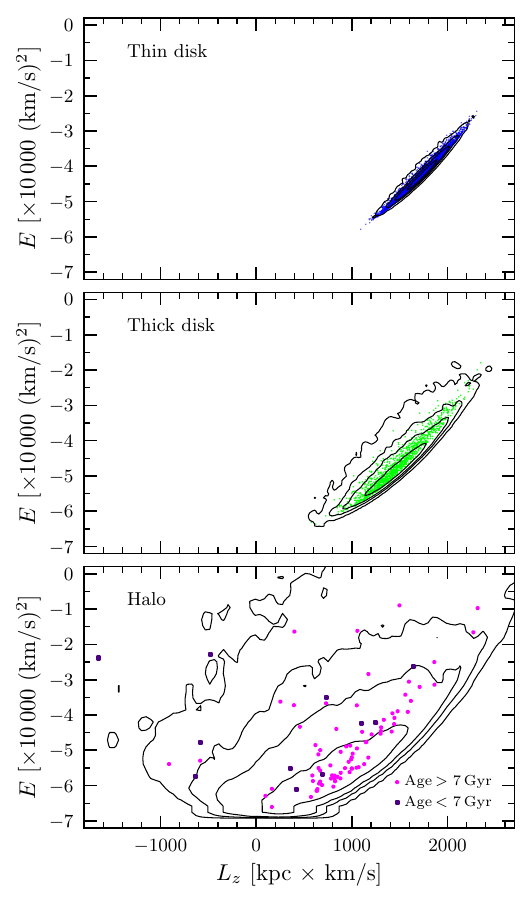}
    \caption{Orbital energy vs. $z$-component of angular momentum. The solid contours represent the 1, 3, and 5\,$\sigma$ levels and the 99th percentile of the simulated sample, while the colored symbols represent the observed white dwarfs. The {\em Gaia} sample classified by \citet{torres2021} as halo white dwarfs is split into two age subgroups.}
    \label{fig:energy_lz}
\end{figure}

Thin-disk stars (top panel of Fig. \ref{fig:ecc_Lz}), for which both synthetic and observed samples appear to be in a reasonable good agreement, present high $L_z$ values and small eccentricities. Both features are strongly related to circular orbits in the disk plane. Thick disk members (middle panel) slightly tend to more eccentric orbits. In this case, when comparing observed and synthetic samples, there is a clear lack of low eccentric thick disk orbits in the former one. As stated in Section \ref{s:obsample}, this fact is attributed to the higher probability of the RF algorithm classifying low-speed circular orbits as thin-disk stars. Finally, the distribution of halo objects (bottom panel) results in a broader range of angular momenta for the observed sample compared to the synthetic sample. There are even some observed halo members that exhibit a negative $L_z$, indicating that they rotate in the opposite direction with respect to the Galactic rotation (retrograde orbits), with very low angular momenta.

We performed a deeper analysis of the distributions of $R_{\rm g}$, $e$, and $L_z$ for the halo stars. We divided the simulated sample in three eccentricity groups separated at intervals of $\Delta e = 0.33$ and in two $L_z$ groups, dividing the sample into prograde ($L_z > 0$) and retrograde orbits ($L_z < 0$). The unit area-normalized distributions of $R_{\rm g}$ for each sub-group are shown in Fig. \ref{fig:Rg_e_Lz}.

We note an intuitively clear relation between high $e$ and low $R_g$. Provided that a very elliptic orbit has a lower energy than the corresponding circular orbit, both crossing the solar circle at 8\,kpc, the elliptic orbit will have its apocenter in the solar neighbourhood end; thus, its pericenter will be closer to the Galactic center. For this reason, there will be a large number of stars passing through inner regions. Simulated halo stars that display small orbit eccentricity with $R_{\rm g}$ values that are more concentrated towards the solar circle, where  circular orbits are usual. The correlation between $L_z$ and $R_{\rm g}$, bottom panel in Fig. \ref{fig:Rg_e_Lz}, shows that all the retrograde orbits ($L_z< 0$\,kpc\,\kms) distinctly cluster at $R_{\rm g} < 5$\,kpc. This result could imply that the few observed white dwarfs with retrograde orbits could have acquired their highly eccentric orbits after being scattered from the inner regions of the Milky Way. Nevertheless, also stellar streams that have formed via perturbations due to past encounters of the Milky Way with other satellite galaxies are characterized  by retrograde orbits \citep{helmi2018,myeong2019}. On the other hand, halo stars with prograde orbits ($L_z > 0$\,kpc\,\kms) have a wider distribution of $R_{\rm g}$ that mimics that of the overall subgroup, indicating a variety of possible origins; these would involve interactions attributed to the outer Lindblad resonance of the Galactic bar playing a major role. In this context, it is worth comparing the distribution of orbital energy and $z$-component of angular momentum for the simulated and observed white dwarfs as shown in Fig.\,\ref{fig:energy_lz}. While thin and thick disk members occupy a limited portion of the diagrams, which is related to their more circular orbits as it is shown in Fig.\,\ref{fig:ecc_Lz}, the halo white dwarfs can display a much larger variety of angular momenta and energies. The latter group consists of  89 stars that were classified by \citet{torres2021}, with 11 of them possessing unusually young ages of less than 7\,Gyr. This subgroup of stars appears to possess comparatively larger energies for a given value of $L_z$, placing  most of them beyond the  1\,$\sigma$ contour of Fig.\,\ref{fig:energy_lz}. While there is an extremely very low chance of contamination with the most extreme thick disk stars, those authors suggested that these  11 white dwarfs could possess halo kinematics, despite their young age, due to more eventful lives as binary-star merger products or relics of past mergers of the Milky Way with its satellite galaxies. %

\begin{figure}
    \centering
    \includegraphics[scale = 0.15]{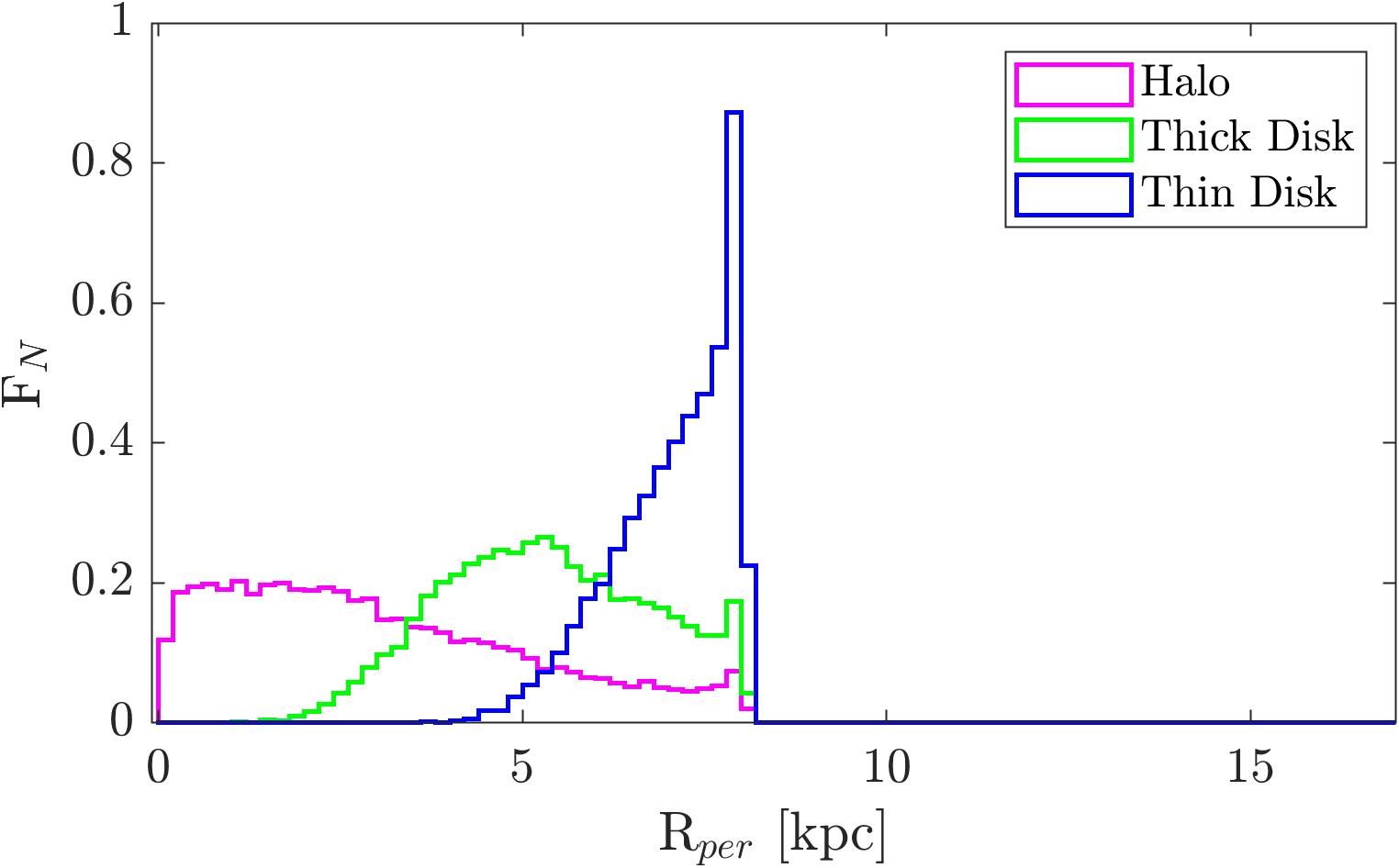}
    \includegraphics[scale = 0.15]{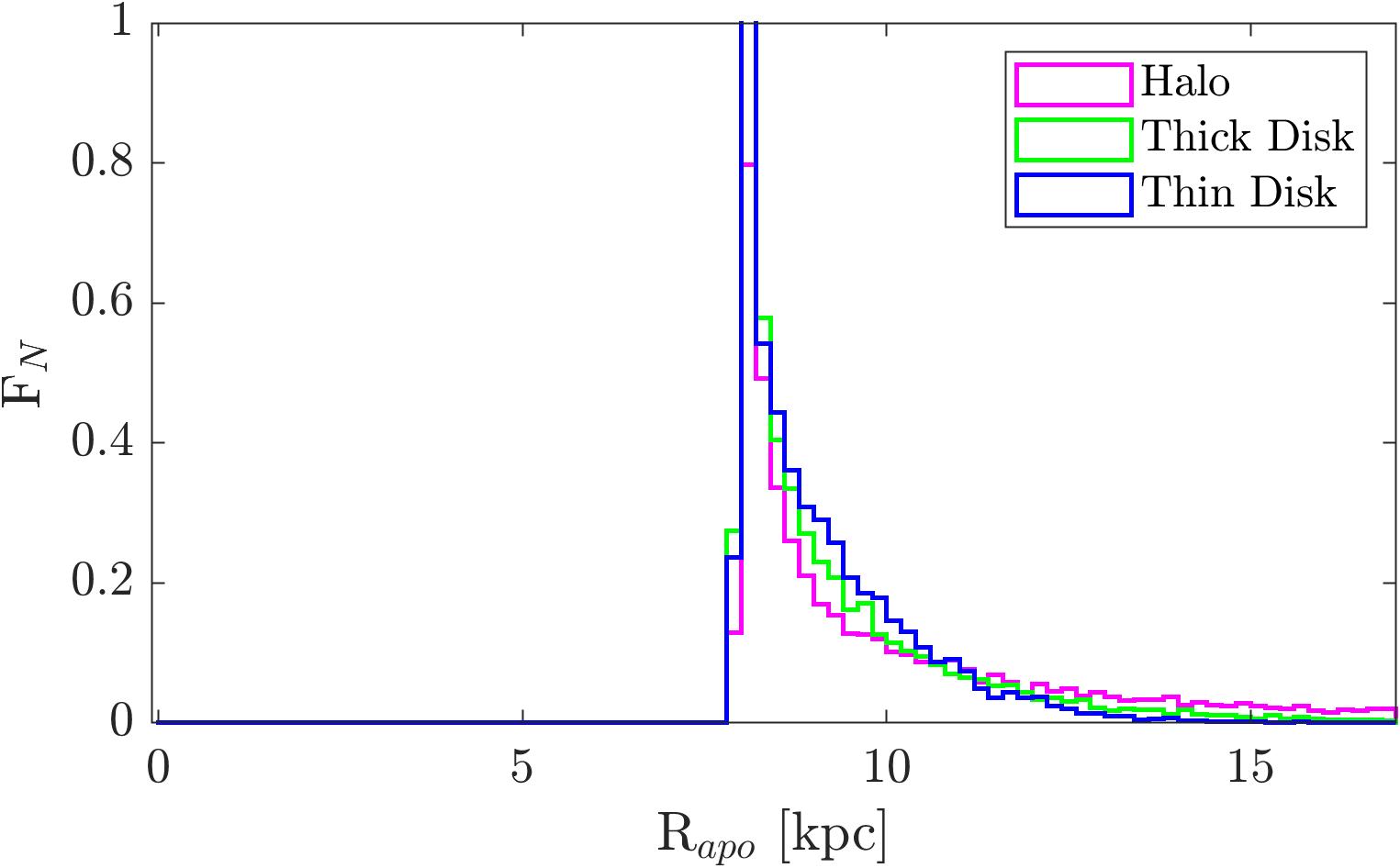}
    \includegraphics[scale = 0.15]{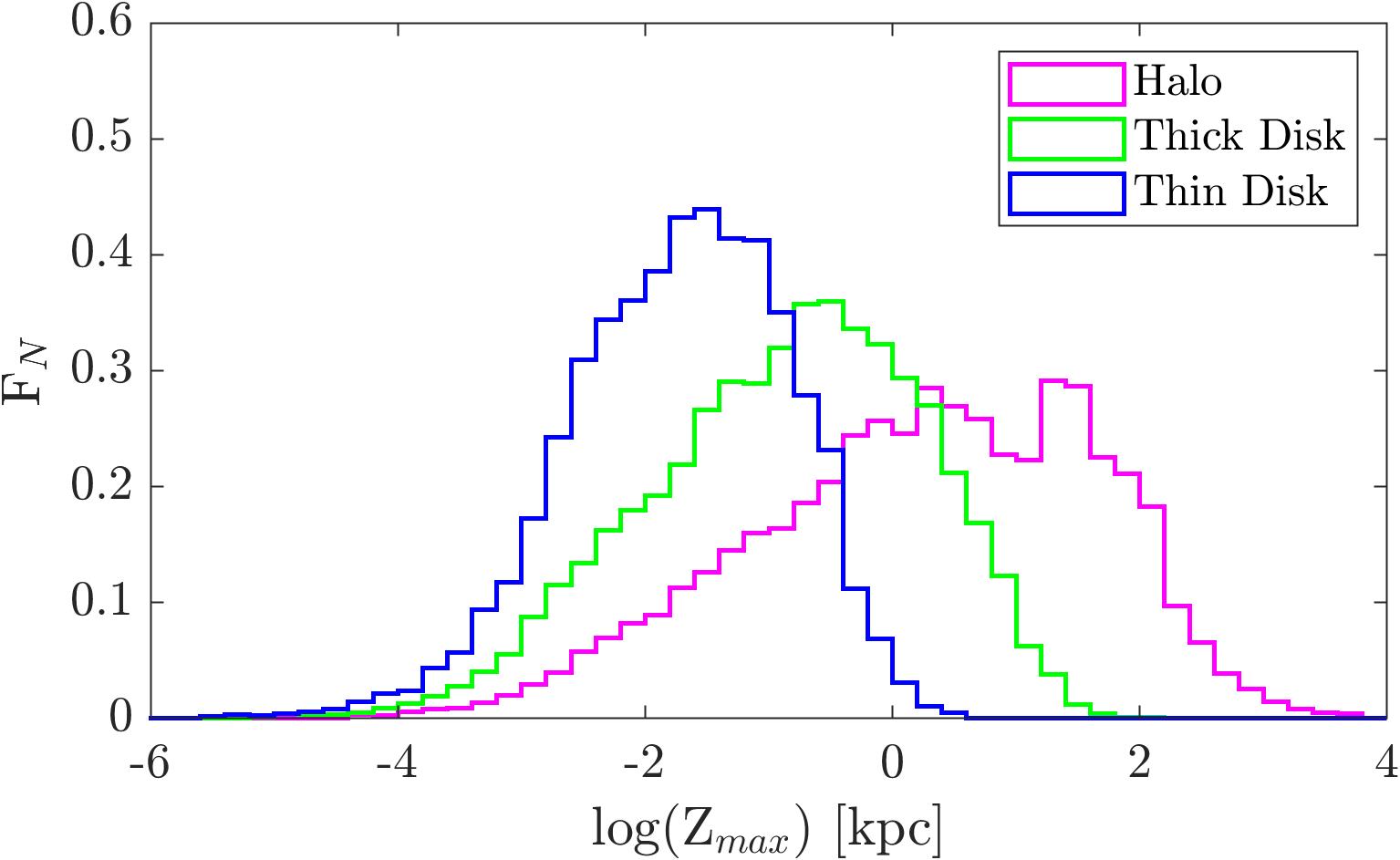}
    \caption{Distribution of pericenters (top pannel), apocenters (middle pannel) and maximum $z$ (bottom pannel) of the synthetic sample. }
    \label{fig:peri_apo_zmax}
\end{figure}

Finally, in Fig.\,\ref{fig:peri_apo_zmax}, we display the distributions of other relevant parameters of kinematic analysis for the simulated samples, which are the pericenter and apocenter radii ($R_{\rm per}$ and $R_{\rm apo}$, respectively), and the maximum vertical displacement ($z_{\rm max}$). By definition, $R_{\rm per}$ and $R_{\rm apo}$ are combined into the eccentricity of an elliptical orbit, thus they show wider distributions with respect to those of $R_{\rm g}$ that is representative of an equivalent circular orbit (Fig\,\ref{fig:Rg} and \ref{fig:Rg_e_Lz}). On the other hand, the $z_{\rm max}$ distributions encode the contribution of the vertical velocity component, $W$, to the orbital motion and are related to the average scale height of the Galactic components. The average values of these quantities for the simulated population are listed in Table\,\ref{table:apo_peri_zmax}.

\begin{table}
\caption{Median values and 16--84 percentile uncertainties for the apocenter, pericenter and maximum $z$ distributions of each galactic component.}
\label{table:apo_peri_zmax}
\centering                         
\begin{tabular}{@{}lccc@{}}
\hline 
      \noalign{\smallskip}
& $R_{\rm peri}$ [kpc] & $R_{\rm apo}$ [kpc] & $z_{\rm max}$ [kpc] \\ 
      \noalign{\smallskip}
\hline 
      \noalign{\smallskip}
Thin disk & $ 7.22_{-1.04}^{+0.67}$ & $ 8.74_{-0.65}^{+1.49}$ & $ 0.19_{-0.11}^{+0.24}$ \\
Thick disk & $ 5.27_{-1.45}^{+1.67}$ & $ 8.68_{-0.60}^{+1.99}$ & $ 0.46_{-0.34}^{+0.84}$ \\
Halo & $ 2.67_{-1.76}^{+2.69}$ & $ 9.33_{-1.17}^{+5.50}$ & $ 1.43_{-1.13}^{+3.88}$\\
      \noalign{\smallskip}
\hline     
\end{tabular}
\end{table}

\subsection{Kinematic effects on the solar neighborhood}
\label{s:effects}

\subsubsection{The solar neighborhood size}
\label{ss:size}

The exact size of the solar neighborhood is an arbitrary value. It depends on which stellar objects we are considering and what the ultimate purpose of the study is. In the case of the white dwarf population, the concept of the solar neighborhood can vary from a few tens to a few hundred parsecs. For instance, there are studies available in the literature on the star formation history up to 40 pc \citep{cukanovaite2023},  the age-metallicity relation up to a few hundred pc \citep{Rebassa2021}, and the white dwarf spectral type-temperature distribution up to 500 pc \citep{Torres2023}.

On the other hand, as we discuss in Section \ref{ss:lifetime}, the solar neighborhood is a transient concept. We now address the question of whether the different volumes that we may consider as the solar neighborhood are equivalent, that is, if the objects that make up this sample are representative of the same region of the Galaxy.

To analyze how the size of the solar neighborhood affects the spatial distribution, we simulated volumes of  40,\,100,\,and 500\,pc and an extra large volume of 1.5\,kpc, assuming in all cases the same velocity distribution as for the 100\,pc sample. In Table \ref{t:rings}, we summarized the two main parameters that characterize the spatial distribution of the white dwarf Galactic components: the width of the ring, $\Delta R$, and the width of the perpendicular layer with respect to the Galactic plane, $\Delta z$. 

The analysis in Table \ref{t:rings} reveals that for solar neighborhood volumes up to 500\,pc, differences in  $\Delta R$ for the Galactic spatial distribution of white dwarfs are minor, while the $\Delta z$ values see a 30--50\,\% increase for the thin- and thick-disk populations, respectively. This implies that current white dwarf surveys up to 500\,pc (e.g., practically 90\% of {\it Gaia} spectra are contained within that volume) are representative of the same putative Galactic white dwarf population.  In other words, the Galactic origin of the vast majority of white dwarf within 40, 100, or 500\,pc from the Sun can be assumed to be the same. Hence, the corresponding star formation histories, luminosity functions, mass distributions, or other ensemble properties extracted from the different samples can be considered to be equivalent.

However, that is not the case for the largest sample analyzed, the 1.5\,kpc volume. In particular, with respect to the disk populations, the Galactic region contributing to a 1.5\,kpc solar neighborhood radius is considerably larger than in previous cases. For instance, the  ring width has  an increased by approximately a factor of 1.3 compared to the 100\,pc sample, while the dispersion in the $z$ coordinate has nearly  doubled. As a result, future analyses of this larger sample should consider incorporating a wider variety of stars, such as those with different metallicities coming from a broader region of the Galaxy. 

 Finally, we note that the halo population does not present a substantial difference in $\Delta R$ for the different neighborhood sizes and it is only characterized by an 30\,\% increase in $\Delta z$, implying that the velocity distribution of inner-halo white dwarfs detected within 100-pc could also be representative for the much larger volume.

\begin{table}
\caption{Spatial distribution of the white dwarf Galactic components as a function of the neighborhood size.}
\label{t:rings}
\centering
\begin{tabular}{lcccc}
\hline
      \noalign{\smallskip}
 Spatial distribution  & \multicolumn{4}{c}{Solar neighborhood size} \\
& 40\,pc & 100\,pc & 500\,pc & 1.5\,kpc \\
      \noalign{\smallskip}
\hline
      \noalign{\smallskip}
Thin disk  $\Delta R$ [kpc] &  2.04 &  2.01 &  2.09 &  2.60 \\
Thin disk $\Delta z$ [kpc] &  0.31 &  0.33 &  0.45 &   0.59 \\
      \noalign{\smallskip}
\hline
      \noalign{\smallskip}
Thick disk  $\Delta R$ [kpc] &  3.49 &  3.40 &  3.45 &  3.67 \\
Thick disk $\Delta z$ [kpc] &  0.78 &  0.80 &  0.96 &  1.40 \\
      \noalign{\smallskip}
\hline
      \noalign{\smallskip}
Halo $\Delta R$ [kpc] &  6.40 &  6.74  &  6.93 &  6.83 \\
Halo $\Delta z$ [kpc] &  2.52 &  2.54 &  2.7 &  3.14 \\
      \noalign{\smallskip}
\hline
\multicolumn{5}{l}{Note: Adopted values correspond to the $1\sigma$ level.}
\end{tabular}
\end{table}
\subsubsection{Effect of null radial velocity assumption}
\label{ss:radial}
We conducted a simulation to assess the impact of assuming null radial velocities on our estimation of the spatial distribution of the solar neighborhood population. In this simulation, we artificially inflate the mean values and dispersions by one-third compared to those adopted in Table \ref{table:vel-gauss}  \citep[see discussion in Sects.\,\ref{s:obsample} and][]{rowell2019}. The results corresponding to the spatial distribution parameters (ring width, $\Delta R$, and $z$-dispersion, $\Delta z$, along with the mean values and standard deviations for the guiding-center radius distributions of each Galactic component) are summarized in Table \ref{table:vrad}. Additionally we selected a sample of 20\,469 FGK-type main-sequence and sub-giant stars within 100\,pc from the Sun, which have reliable {\em Gaia} astrometry and radial velocities (with the first 13 parameter flags of the \verb|flags_gspspec|  string set equal to zero; \citet{recio-blanco2023}). This sample of non-evolved stars occupies the same volume as that of the observed white dwarfs and, thus, it offers a useful comparison for the bias introduced by accounting or not accounting for radial velocities of stars. As for the local white dwarf sample, it is mostly comprised of thin-disk objects, but also some thick-disk and halo stars are present. As for the white dwarfs, we analyzed these stars  by integrating their orbits  but,  in this case, we also accounted for their measured radial velocities.  The results are listed at the last row of Table\,\ref{table:vrad}.

From the comparison with the previously obtained results (see Tables \ref{table:Rg} and \ref{t:rings} for the 100\,pc simulation),  we deduced that inflating the space velocity by one-third increases the volume occupied by the orbits of thin-disk white dwarfs by 30--35\,\%, as quantified by $\Delta R$ and $\Delta z$, and reduces by 2\% the mean $R_g$ with a 40\,\% larger $\sigma_{R_g}$.  A similar result, but on the order of 10--15\,\%, is found when considering the main sequence and sub-giant sample, which naturally incorporates measurements of the radial velocity. However,  the inclusion of radial velocities would introduce a much more significant change in the volume occupied by thick-disk and halo white dwarfs, which appear to double in size for the latter sub-sample that is characterized by more extreme kinematics. The greater reduction in the guiding radius of a halo white dwarf is thus accompanied by an increment of the eccentricity of their orbits. A consequence of these changes, there is a  $\approx 30$\,\% reduction of the solar neighborhood lifetime with respect to the results of Section\,\ref{ss:lifetime}, corresponding to 2.47, 1.04, and 0.48\,Myr for the thin-disk, thick-disk, and the halo, respectively.

In summary, we conclude that the inclusion of an estimate of the radial component of the velocity of white dwarfs  is currently rather uncertain. However, if their kinematics is similar to that of main sequence and sub-giants belonging to the same sub-populations, our results will be valid within a systematic 15--30\,\% uncertainty.

\begin{table}
\caption{Spatial distribution parameters, mean values and standard deviations for the guiding-center radius distributions of each Galactic component when a radial velocity estimate is taken into account.}
\label{table:vrad}
\centering                         
\begin{tabular}{@{}lcccc@{}}
\hline 
      \noalign{\smallskip}
& $\Delta R$\,(kpc) & $\Delta z$\,(kpc)  &$\langle$R$_g$$\rangle$\,(kpc)  & $\sigma_{R_g}$\,(kpc) \\ 
      \noalign{\smallskip}
\hline 
      \noalign{\smallskip}
Thin disk &  2.62 &  0.45 &  7.73 &  0.97 \\
Thick disk &  4.48 &  1.23 &  6.10 &  1.67 \\
Halo &  11.12 &  4.87 &  3.65 &  2.68 \\
MS/SG &  2.21 &  0.35 &  7.55 &  1.00 \\
      \noalign{\smallskip}
\hline
\multicolumn{5}{l}{Note: A comparison with main sequence (MS) and sub-}\\
\multicolumn{5}{l}{giant (SG) stars within 100\,pc from the Sun is included.}
\end{tabular}
\end{table}

\subsubsection{Selection criteria effects}
\label{ss:selection}
\begin{table}
\caption{Simulated stars that are either observable or not observable by {\em Gaia}.}
\label{table:Gaiaobs}
\centering                         
\begin{tabular}{@{}ccc@{}}
\hline 
      \noalign{\smallskip}
Population & Total & Observable (fraction) \\ 
      \noalign{\smallskip}
\hline 
      \noalign{\smallskip}
Thin disk & 15027 & 11949 (79.46\%) \\
Thick disk & 13861 & 6753 (48.72\%) \\
Halo & 15094 & 6071 (40.22\%) \\
      \noalign{\smallskip}
\hline     
\end{tabular}
\end{table}
\begin{figure}
    \centering
    \includegraphics[scale = 0.15]{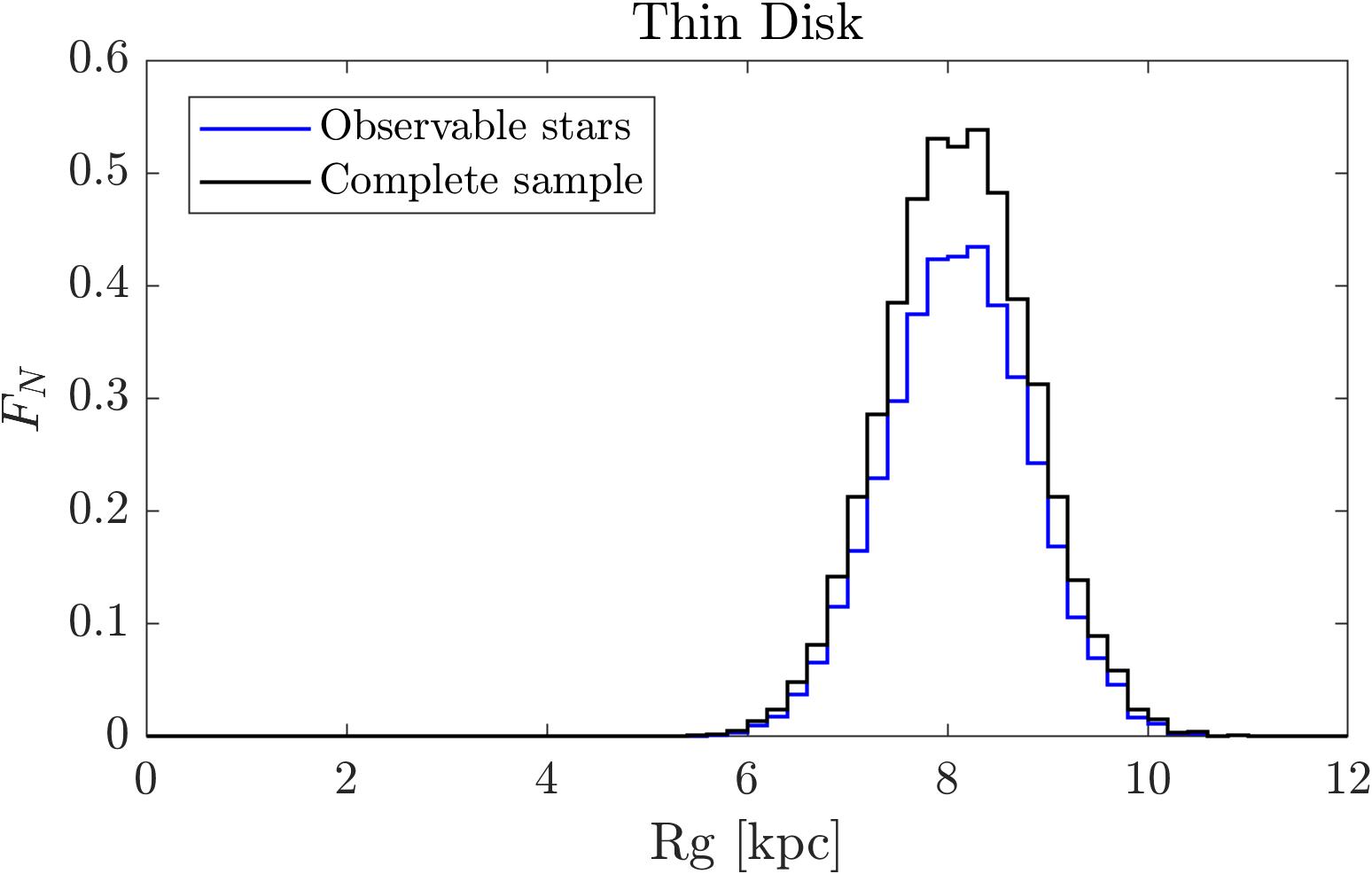}
    \includegraphics[scale = 0.15]{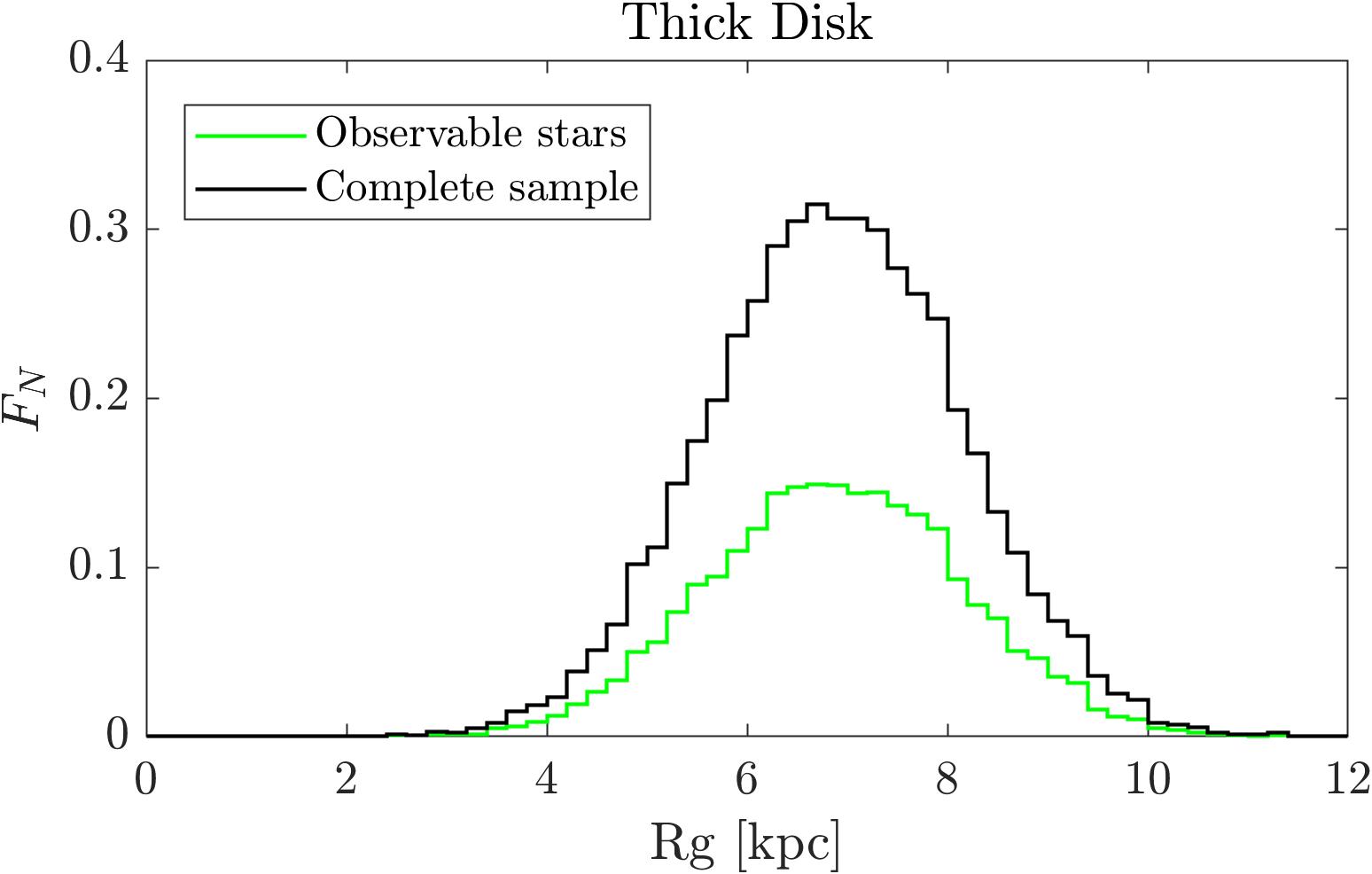}
    \includegraphics[scale = 0.15]{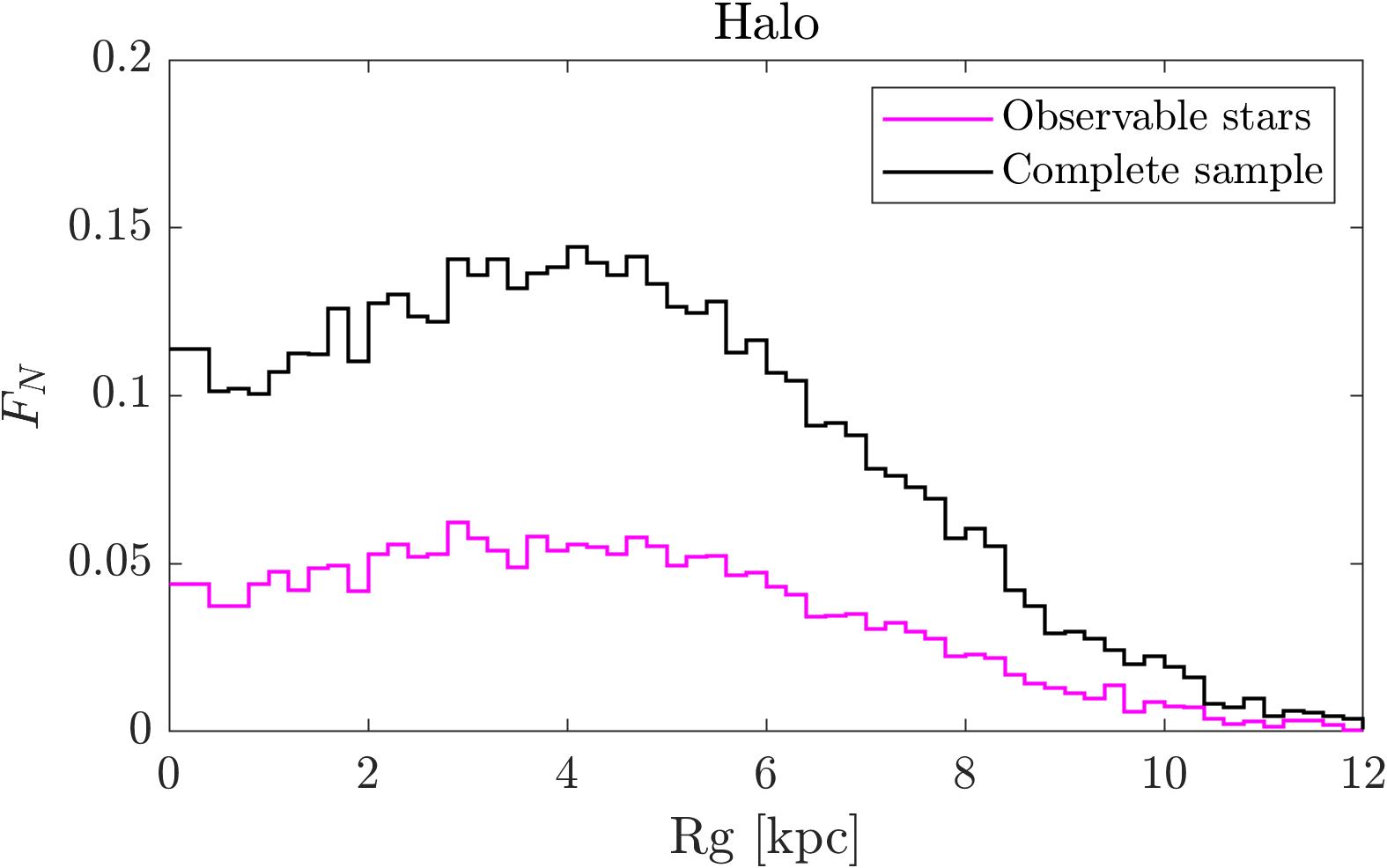}
    \caption{Normalized guiding-center radius distribution of white dwarfs that would be observed by {\it Gaia} (colored histograms) and the complete sample (empty histograms).}
    \label{fig:Rg_gyn}
\end{figure}

We have stated that our {\it Gaia} 100 pc sample can be considered  effectively complete. However, it is well known that the overall completeness of the sample rapidly decreases for red colors and faint magnitudes \citep[see, e.g., Fig. 5 from][]{JE2018,Gentile-Fusillo2021}. It is to be expected that this bias will have an effect on the average values of the observed velocities, as it is more likely that older objects will be excluded from the observed sample. 

To check the effect of the selection criteria imposed by {\it Gaia}, we associate to each synthetic star a mass and a total age according to the Galactic models described in \cite{Torres2019}. By using appropriate cooling sequences from the set of evolutionary models of the La Plata group \citep{althaus2013,camisassa2016,camisassa2019}, we can assign \emph{Gaia} magnitudes to each simulated white dwarf. Furthermore, our MC simulator is able to emulate the {\em Gaia} performance. Finally, we introduce a selection cut for those objects with astrometric errors and photometric flux errors larger than 10\% and an apparent magnitude in {\em Gaia} G filter below 21\,mag. Thus, the synthetic objects in our initial sample are split into those that pass {\it Gaia}'s selection criteria, and thus would be observable, and those that are excluded.

The results listed in Table \ref{table:Gaiaobs} show that while the majority of the thin-disk objects can be observed by {\em Gaia}, a much larger fraction of halo objects can not. Thick disk objects are just in the middle, with about 50\% of them not being observable by {\em Gaia}.

The question that we now rise is if the non-observed white dwarfs contain meaningful information that observed white dwarfs do not show. To address this issue, the simulated sample has been divided and studied separately. The observed stars were compared to the complete sample in the previous sections. If the guiding-center radius distribution is considered, the panels in Figure \ref{fig:Rg_gyn} show that observable objects (colored histograms) reproduce the same distribution as for the complete sample (empty histograms), despite a reduced fraction previously mentioned (see Table \ref{table:Gaiaobs}).  A similar effect is observed in the rest of orbital parameters, such as the $z$ angular momentum, eccentricity or energy. Consequently, we conclude that, except for a statistical reduction in the number of observable objects,  {\it Gaia} selection effects have not induce a kinematic bias in the various Galactic white dwarf populations within 100\,pc.

\section{Conclusions}

We have studied the kinematics and the orbital parameters of a synthetic white dwarf population, which was modeled from the observed properties of local white dwarfs within 100\,pc away from the Sun. The simulated population was divided into thin- and thick-disk, and halo components to compare with the results of the Random Forest classification of the \emph{Gaia} sample \citep{Torres2019}. 

By numerically integrating the Galactic orbits of simulated and observed stars, we have showed that the solar neighborhood is a temporary concept, with an average lifetime of  0.6--3.3 Myr, depending on the average speed of the considered population. Nevertheless, most of the local white dwarfs that belong to the thin disk seem to originate from a relatively small annulus of  about $\pm 1$\,kpc in width that is centered on the LSR; 
 this implies that most of of the thin-disk white dwarfs have progenitors that are chemically
and evolutionarily similar. The analysis of the average orbits and the distributions of guiding-center radii for thick disk and halo stars is characterized by a much broader coverage of the available space parameters, relating that to the radial mixing of stars. It is well established that these effects are likely due to the interaction with the Galactic bar and/or spiral arms potentials. We note a significant correlation among eccentric-and-retrograde and less-eccentric-and-prograde orbits of halo stars and their distribution of guiding-center radii. This possibly suggests a different origin either from accreted satellite galaxies or the existence of an in-situ stellar population in the inner halo. 

By simulating possible detection biases, which are either due to the \emph{Gaia} detection sensitivity or to the kinematic classification of the observed sample, we show that their importance is relatively small. In fact, the good agreement between simulated and observed crossing times of the solar neighborhood are indicative that there is small contamination among kinematic populations. The largest divergence between lifetimes for the observed and simulated samples is noted for the thick disk population, for which we do not model the more complex kinematic signature of stellar streams. Moreover, the \emph{Gaia} detection bias does not affect the overall parameter space, which allowed us a reliable coverage of the simulated distributions. 

The bias introduced by the known lack of accurate radial velocities for white dwarfs was previously discussed by \citet{Torres2019}  and \citet{rowell2019}, could account for systematic differences that could be rather large for thick disk and halo stars. However, the comparison with local main-sequence and sub-giant stars suggests that the systematic uncertainties affecting the estimates of the solar neighborhood lifetime and space volume occupied by the Galactic orbits of the local sample should be in the range of 15--30\,\%. Nevertheless,  we stress the importance of measuring accurate radial velocities of white dwarfs. We expect that this issue will be partly resolved in the next five to ten years thanks to the spectroscopic observations of the majority of \emph{Gaia} white dwarfs by the WEAVE \citep{weave} and 4MOST \citep{4most} surveys, which will additionally provide metal abundances for a sub-sample of white dwarfs in wide binary systems with a non-degenerate companion \citep{wdbs}. Such larger datasets will enable a more detailed analysis of the existing correlations among age, kinematics, and metallicity, which we recently studied for magnitude-limited samples of white dwarfs \citep[e.g.,][]{Rebassa2021,Raddi2022} and are known to be related to the dynamical evolution of our own Galaxy.

\begin{acknowledgements}
 We thank the anonymous referee for constructive comments that helped to improve the manuscript.
 
 RR acknowledges support from Grant RYC2021-030837-I funded by MCIN/AEI/ 10.13039/501100011033 and by “European Union NextGeneration EU/PRTR”. This work was partially supported by the AGAUR/Generalitat de Catalunya grant SGR-386/2021 and by the Spanish MINECO grant PID2020-117252GB-I00. 
 
 This work has made use of data from the European Space Agency (ESA) mission
{\it Gaia} (\url{https://www.cosmos.esa.int/gaia}), processed by the {\it Gaia}
Data Processing and Analysis Consortium (DPAC,
\url{https://www.cosmos.esa.int/web/gaia/dpac/consortium}). Funding for the DPAC
has been provided by national institutions, in particular the institutions
participating in the {\it Gaia} Multilateral Agreement.
\end{acknowledgements}

\bibliographystyle{aa}
\bibliography{KOWD}

\end{document}